\documentstyle[preprint, aps]{revtex}

\def\mulref{\cite {artemenko,nazarov1,yip,nazarov2,volkov1,golubov}}

\begin{document}
\draft

\title{Reentrance effect in normal-metal/superconducting hybrid loops}

\author{C.-J. Chien and V. Chandrasekhar}
\address{Department of Physics and Astronomy, Northwestern University, Evanston,
Illinois 60208}

\date{\today}

\maketitle

\begin{abstract}

We have measured the transport properties of two mesoscopic hybrid loops composed
of a normal-metal arm and a superconducting arm.  The samples differed in the
transmittance of the normal/superconducting interfaces.  While the low
transmittance sample showed monotonic behavior in the low temperature resistance,
magnetoresistance and differential resistance, the high transmittance sample
showed reentrant behavior in all three measurements.  This reentrant behavior is
due to coherent Andreev reflection at the normal/superconducting interfaces.  We
compare the reentrance effect for the three different measurements and discuss the
results based on the theory of quasiclassical Green's functions.
\end{abstract}

\widetext

\section{Introduction}
\label{Intro} The resistance of a normal metal (N) in contact with a
superconductor (S) is modified in the vicinity of the N-S  interface, a phenomenon
well known as the superconducting proximity effect
\cite{deu}.  The microscopic mechanism of the proximity effect is Andreev
reflection at the N-S interface:  An electron in the normal metal
with energy less than the gap of the superconductor is Andreev-reflected as a hole,
with the concurrent generation of a Cooper pair in the superconductor
\cite{andreev}. The consequence of this mechanism is the existence of a
superconducting correlation in the normal metal.  Recently, using a theory based
on quasiclassical nonequilibrium Green's functions, it was shown that the
superconducting correlation is expected to decay over a length
$\xi(\epsilon$)=\(\sqrt{\hbar D_N/\epsilon}\), where
$\epsilon$ is the energy of the electron and $D_N$ the diffusion constant of
electrons in the normal metal \mulref.  The surprising result was predicted that
the resistance of the normal metal returns to its normal state value at zero
temperature and energy, the so-called reentrance effect \mulref.

The physical manifestation of this reentrant behavior can be seen in the 
transport properties which are described by an effective diffusion constant
$D(\epsilon,x)$, a quantity dependent on the energy of the electron $\epsilon$ and
the position $x$ \cite {nazarov2,volkov1,golubov}.  $D(\epsilon,x)$ coincides
with its normal state value at zero energy, increases and reaches a maximum at an
intermediate energy of the order of the Thouless energy $E_{c}=\hbar D_N/L^{2}$,
and coincides with its normal state value again at higher energies ($L$ here is
the length of the normal metal).  As a function of temperature, the resistance of
a diffusive normal metal adjacent to a superconductor shows a minimum at a
temperature
$T$ of the order of $E_{c}/k_{B}$ (where $k_{B}$ is the Boltzmann constant), and
regains its normal state value as $T
\rightarrow 0$.  A similar minimum is expected in the differential resistance of
the normal metal as a function of dc voltage $V$.  If the diffusive normal metal
is connected to two superconductors with different phases, the resistance is
expected to oscillate as a function of the phase difference between the
superconductors, which can be modulated by the application of a magnetic field. 
The amplitude of the magnetoresistance oscillations shows a maximum when $T$ is of
the order of $E_{c}/k_{B}$ and vanishes again as $T \rightarrow 0$  (neglecting
electron-electron interaction in the normal metal) \cite {nazarov2,volkov1,golubov}.  All this is strictly valid only for high interface transmittances; if
the interface transmittance is low, the probability of Andreev reflection is
correspondingly reduced, and the reentrant behavior in the transport properties is
shifted to lower energy scales, and may disappear entirely \cite {yip,volkov2}.

Several groups have reported observing this reentrance effect in normal
metal-superconductor (N-S) or semiconductor-superconductor (Sm-S) structures.   In
N-S structures, Courtois \emph{et al.} \cite {courtois} reported observing
magnetoresistance oscillations in a normal Cu loop with two superconducting Al
islands on either side. No reentrant behavior was observed, however, possibly
because it was masked by the Josephson coupling between the superconducting
islands.  Charlat \emph{et al.} \cite {charlat} observed the reentrance effect as
a function of both temperature and voltage in a Cu loop adjacent to a small
superconducting Al island.  In a different geometry, Petrashov \emph{et al.} 
\cite {petrashov1} observed the reentrance effect in the amplitude of
magnetoresistance oscillations in an Andreev interferometer.  In Sm-S structures,
den Hartog \emph{et al.} \cite {denHartog1,denHartog2} have reported observing
reentrant behavior in a geometry where a diffusive two dimensional electron gas is
coupled to a superconductor to form a loop.  In these experiments, den Hartog
\emph{et al.} observed reentrant behavior not only in the resistance as a function
of dc voltage, but also in the amplitude of magnetoresistance oscillations as a
function of dc voltage. Toyoda \emph{et al.} \cite {toyoda} observed reentrance in
the magnetoresistance oscillations of a two dimensional electron gas connected to
a superconducting loop.  While the results of these experiments qualitatively
agree with the theory, quantitative comparison with theory is still not
satisfactory, especially at high energies or temperatures close to
$T_{c}$, where quantitative predictions are difficult to obtain.

In this paper, we report detailed transport measurements on two N-S devices as a
function of temperature $T$, magnetic field $H$ and dc voltage bias $V$.  Both
devices are in the form of square loops, with one arm of the loop fabricated from
a normal metal (Ag or Au) and the remaining three arms from a superconductor (Al). 
The primary difference between the two samples is in the transmittance of the N-S
interfaces:  One sample (sample A) has low interface transmittances ($R_{b}/R_{N}>
1$, where
$R_{b}$ is the interface resistance and $R_{N}$ is resistance of the normal
metal), while the other sample (sample B) has high interface transmittances
($R_{b}/R_{N}\ll 1$).  Both samples show a strong temperature dependent resistance
$R(T)$, large oscillations in the magnetoresistance $R(H)$, and a differential
resistance $dV/dI$ which is a function of $V$ below the critical temperature
$T_{c}$ of the superconductor.  While sample A shows monotonic behavior in all
three measurements, sample B shows reentrant behavior in $R(T)$, $dV/dI(V)$, and
amplitude of oscillations in $R(H)$ as a function of $T$.  The temperature and
energy scales for this reentrant behavior are in qualitative agreement with recent
theories on Andreev reflection in mesoscopic N-S devices \mulref, although
detailed quantitative agreement is still lacking.

\section{SAMPLE FABRICATION AND MEASUREMENT}
\label{Samp}

An electron beam micrograph of sample A is shown in Fig. 1(a), and schematics of
the two samples A and B are shown in Fig. 1(b).  The samples were fabricated by a
conventional multi-level electron-beam lithography process.  The normal metal was
deposited first.  After a second level of e-beam lithography, the normal metal
surface was cleaned by a dc $Ar^{+}$  etch and the superconductor (Al) was
deposited without breaking vacuum in order to ensure good contacts between N and
S.  Au was used as the normal metal for sample A, while Ag was used for sample B. 
Control Ag and Al wires were coevaporated with sample B in order to calibrate film
properties.  The relevant film parameters are as follows: Au/Ag thickness $\sim$
28 nm, Al thickness $\sim$ 37 nm, wire linewidth
$\sim 0.1-0.14$ $\mu m$, normal metal (Ag) coherence length  $\xi_{N}(T) \sim
0.23$ $\mu m/ \sqrt{T}$, superconducting coherence length $\xi_{Al}(T=0)=0.31$
$\mu m$, and electron phase coherence length $L_{\varphi} = 0.9$ $\mu m$ at $T$=30
mK \cite {lphinote}.  The area of the N-S interfaces was approximately 0.15x0.15
$\mu m^{2}$.  The samples were measured in a dilution fridge between 30 mK and 1.5
K using a four-terminal ac resistance bridge, with ac excitations in the range of
10-100 nA, small enough to avoid self-heating. The four terminal measurement
configuration is shown in Fig 1(b).  For the $dV/dI$ measurements, the dc current
was applied through the same leads as the ac current.  Aside from the transparency
of the interfaces, the major difference between the two samples is the greater
length of the normal metal arms beyond the loop in sample A in comparison to
sample B (see Fig. 1(b)).

\section{EXPERIMENTAL RESULTS}
\label{Exp}

Figure 2 shows the resistance of both samples as a function of temperature. The
first difference noticeable  between the two samples can be seen near $T_{c}$. 
For sample B, there is a sharp drop in resistance as the sample is  cooled through
$T_{c}$.  If the N-S interface resistances were negligible, one might expect the
superconducting arm to  short out the normal arm of the loop, resulting in a
decrease in resistance corresponding to the normal state  resistance of the loop
alone.  This is indeed the change in resistance we observe for sample B within our
experimental  error, based on the measured resistivities of the normal metal and
superconductor.  Thus, the interface in sample B  appears to be highly
transparent.  Since the resistance of sample A does not show a sharp drop at
$T_{c}$, but only a gradual  and small decrease as the temperature is lowered, we
conclude that the interface transmittances for this sample are small.   We can
also estimate the barrier resistance from the resistivities of the Ag, Au and Al
films along with the measured  resistance of the samples.  Based on these
measurements, the resistance of each N-S interface is $\sim$ 25
$\Omega$ for sample A  and $<$ 0.5 $\Omega$ for sample B. 

The second major difference between the two samples is seen in the low temperature
behavior.  When the temperature is  reduced from $T$ = 1.2 K to 30 mK, the
resistance of sample A decreases monotonically.  Sample B, on the other hand,
eventually shows an \textit{increase} in the resistance, resulting in a minimum in
the resistance at $\sim$ 520 mK.  This is similar to the behavior observed by
Charlat
\emph{et al.} \cite {charlat} in their Cu/Al loops, and was attributed by them to
the anomalous proximity effect in the Cu loop induced by the Al island.  The
temperature at which the resistance minimum
$R_{min}$ occurs is given approximately by the temperature at which
$\xi_{N}(\epsilon=k_{B}T)$ is comparable to length of the relevant normal region. 
For sample B, the normal regions that contribute to the low temperature zero bias
resistance are the small normal arms outside the loop, which have lengths of 0.15
and 0.55 $\mu m$ respectively, since the loop itself has zero resistance below
$T_{c}$.  The resistance at the lowest temperature is 5.9 $\Omega$ which
corresponds to the resistance of the normal side branches at this temperature as
noted above.  Due to its longer length, the contribution from the 0.55 $\mu$m arm
should dominate the electrical transport.  The temperature corresponding to the
Thouless energy for this arm is $T = E_{c}$/$k_{B}$ $\sim$ 170 mK, a factor of 3
lower than the measured temperature of
$\sim$ 520 mK.  However, it should be noted that for a normal wire with one end
connected to a superconducting reservoir and the other end to a normal reservoir,
the minimum is expected to occur at
$T$ $\sim$ 5$E_{c}$/$k_{B}$ [4,6]. For sample A, the absence of reentrant behavior
is consistent with the fact the interface transmittances in this sample
are small.  In addition, the normal side arms are long, and would not be expected
to show reentrant behavior in our temperature range.  If one considers the normal
region in the immediate vicinity of the low transmission interface as a highly
disordered conductor with a very low diffusion coefficient D, the relevant
$\xi_{N}$(T) is very short, and hence the reentrant behavior is pushed to much
lower energies and temperatures \cite {yip,denHartog1}.  

A similar difference between the two samples can be observed by examining the
magnetoresistance oscillations as a function of temperature. Figure 3(a) shows the
magnetoresistance of sample A at a few temperatures below 1 K.  Oscillations of a
period corresponding to a flux $h/2e$ through the loop are observed which persist
up to the critical temperature $T_{c}$ of the superconductor, and whose amplitude
at the lowest temperatures is much larger than $e^{2}/h$ (in terms of
conductance).  Similar oscillations are seen in sample B (Fig. 3(b)).  The
presence of magnetoresistance oscillations points to the existence of a quantum
interference effect involving the doubly-connected loop.  The large amplitude of
these oscillations rules out the possibility of their being due to a normal metal
quantum interference effect such as weak localization or conductance fluctuations,
whose amplitude is typically
$\sim$
$e^{2}$/h, and points to a coherent interference phenomenon involving charge
carriers in the normal and superconducting arms of the loop \cite
{deVegvar,pothier,dimoulas,petrashov2}.
      
Figure 3(c) shows the amplitude of the magnetoresistance oscillations for the two
samples as a function of temperature.  The amplitude is determined by calculating
the power in the Fourier transform in the inverse field range corresponding to the
area of the loop, over the field range +/- 25 mT for sample A and +/- 20 mT for
sample B.  While the oscillation amplitude in sample A shows a monotonic increase
as the temperature is decreased, the amplitude of the oscillations for sample B
displays reentrant behavior with a maximum at a temperature of $\sim$ 200 mK. 
Since the oscillations arise from interference effects around the loop, one might
expect that the amplitude of the oscillations would be determined by the ratio of
$\xi_{N}(\epsilon = k_{B}T$) to half the length $L$ of the normal arm, which is
$\sim 1.1$ $\mu m$.  At $T \sim$170 mK, $2\xi_{N}(\epsilon = k_{B}T) = L$.  This is
in good agreement with the temperature at which we observe the amplitude
maximum.   For sample A, no such maximum is observed, even though the film
parameters for the two samples are similar.  This again is a consequence of the
low N-S interface transparencies in this sample.  At higher temperatures, both
samples show a temperature dependence which is well described by a function of the
form $exp[-\alpha L/\xi_{N}(T)]$, as can be seen in Fig. 3(c).  This is in contrast
to the results of Courtois \emph{et al.} \cite {courtois}, where the
magnetoresistance oscillations were seen to decay as a power law in temperature. 
For comparison, we also show the best fit to the power law dependence found in Ref.
\cite {courtois}, which does not describe the data well.  This difference may
arise from the difference in the geometry of the samples in the two experiments.

The differential resistance $dV/dI$ as a function of $V$ of sample A and sample B
also show differences which are consistent with the difference in the quality of
their interfaces.  Figure 4 shows $dV/dI$ as a function of $V$ for both samples at
low temperature and bias.  Sample B again shows reentrant behavior, with a
resistance minimum at a bias voltage of $\sim$ 7.25 $\mu V$.  As in the temperature
dependent resistance, only the two normal side arms are expected to contribute at
low dc bias.  $E_{c}$ for the longer arm is 15 $\mu eV$, and hence the voltage at
the resistance minimum is
\textit{smaller} than expected by approximately factor of two.  This should be
contrasted with the temperature dependence of this sample which was discussed
earlier, where the temperature at which the minimum in resistance was observed was
\textit{larger} than $E_{c}/k_{B}$ by a factor of three.  This discrepancy will be
discussed later when we attempt to compare these data with the quasiclassical
Green's function theory.

In contrast to sample B, sample A shows only a gradual increase in resistance at
zero field with voltage, consistent with the behavior seen in the temperature
dependence.  At a finite magnetic field of 225 gauss, the curvature of the peak
changes, and $dV/dI$ as a function of $V$ shows what appears to be reentrant
behavior as a function of voltage, similar to that of sample B.  However, this
change in the curvature is not due to reentrance, which can be seen by examining
the temperature dependence of the sample in a finite magnetic field.  Figure 5(a)
shows the temperature dependent resistance of sample A at four different magnetic
fields corresponding to 0, 1/2, 1 and 3/2 flux quanta $h/2e$ through the area of
the loop.  Although the curves for half-integral flux quanta are different from
those for integral flux quanta, no reentrant behavior is observed.  We believe
instead that the change is curvature is similar to the zero bias anomaly behavior
observed by Kastalsky \emph{et al.} \cite {kastalsky}, which was explained by van
Wees
\emph{et al.} \cite {vanWees} as arising from suppression of coherent multiple
Andreev reflections by a magnetic field.  Figure 5(b) shows similar data for
sample B, where curves for both integral and half-integral flux quanta show clear
reentrant behavior.  At half-integral flux quanta, there is a small increase in
the temperature $T_{min}$ at which the resistance minimum occurs, but the curve
for zero and one flux quantum are almost the same.  This is in contrast to the
results of Charlat
\emph{et al.} \cite {charlat}, who saw a monotonic increase in $T_{min}$ as the
magnetic field was increased which they attributed to the field dependence of the
electron phase coherence length
$L_{\varphi}$.  At finite magnetic field, $L_{\varphi}$ is shorter
than at zero field \cite{washburn}.  Since $L_{\varphi}$ defines the cutoff length
for coherent Andreev reflection, $L_{\varphi}$ corresponds to the effective length
of the sample, and hence the minimum in resistance as a function of temperature
would move to higher temperatures as a function of magnetic field.  This is
clearly not seen in our samples. 

\section{Discussion}
\subsection{Quasiclassical Green's function model}

We shall now attempt a quantitative description of the temperature and voltage
dependences of sample B using the quasiclassical Green's function theory.  
Our analysis is based on solving the Usadel equation \cite {usadel} for the
parametrized pair correlation function
$\theta (\epsilon,x)$ in the normal metal \mulref.  $\theta(\epsilon, x)$ is a
function of the energy $\epsilon$ and position $x$.  Assuming that the electron
phase coherence length $L_{\varphi}$ is much longer than the length of the normal
metal, the Usadel equation can be written in the simplified form
\begin{equation}
\frac{\partial^{2}\theta(\epsilon, x)}{\partial x^{2}} + 2i\epsilon
sin\theta(\epsilon, x) = 0
\label{eq:usadel}   
\end{equation} 
The current is then given by the equation developed in Ref. \cite
{golubov} (for the special case of a perfect interface):
\begin{equation} 
I(V, T) = \frac{1}{2R_{N}}\
\int_{0}^{\infty}d\epsilon
\left[ tanh
\left(
\frac{\epsilon+eV}{2k_{B}T}\right)-tanh
\left(\frac{\epsilon-eV}{2k_{B}T}\right)\right]D(\epsilon)\
\label{eq:IV}
\end{equation} where $D(\epsilon)$ is the energy dependent diffusion coefficient
in the normal metal which is given in terms of
$\theta ( \epsilon , x)$ by
\begin{equation} 
D(\epsilon) = \frac{1}{\frac{1}{L}\int_{0}^{L}dx
sech^{2}[Im\theta(\epsilon,x)]}\
\label{eq:diffusion}
\end{equation} 
The conductance is obtained from Eq.\ (\ref{eq:IV}) by taking the
derivative with respect to the voltage $V$, $G(V, T) =dI/dV$.  At zero bias, the
calculation is simplified and the following formula is obtained for the resistance
$R(T)$: 
\begin{equation} 
R(T) =
R_{N}\left[\int_{0}^{\infty}\frac{d\epsilon}{2k_{B}Tcosh^{2}\frac{\epsilon}{2k_{B}T}}\frac{1}{\frac{1}{L}\int_{0}^{L}dx
sech^{2}[Im\theta(\epsilon,x)]}\right]^{-1}\ 
\label{eq:RT}
\end{equation} 
Equation\ (\ref{eq:usadel}) must be solved subject to the appropriate
boundary conditions, which are usually specified at the N and S reservoirs
\cite {yip,nazarov2,golubov}.  At a N reservior, $\theta(\epsilon, x)=0$. At a S
reservoir, 
\begin{mathletters}
\label{boundequations}
\begin{equation}
\theta(\epsilon, x) = \frac{\pi}{2} + i \frac{1}{2} ln \left[\frac{\Delta +
\epsilon}{\Delta - \epsilon} \right]
\label{boundequa}
\end{equation}
for $\epsilon < \Delta$, and
\begin{equation}
\theta(\epsilon, x) = i  \frac{1}{2} ln \left[\frac{\epsilon + \Delta}{\epsilon -
\Delta} \right] 
\label{boundequb}
\end{equation}
for $\epsilon > \Delta$, where
$\Delta$ is the superconducting energy gap.  At the N-S interface,
\begin{equation}
\sigma_{N,S}S \left[ \frac{\partial\theta(\epsilon, x)}{\partial x} \right] =
G_{b}sin[\theta_{s}(L,\epsilon)-\theta_{N}(L,\epsilon)]
\label{boundequc}
\end{equation} where $\sigma_{N,S}$ is the conductivity of the interface, $S$ is
the cross section of the wire and $G_{b}$ is the conductance of the interface. 
Finally, at a node where two or more normal wires intersect, the boundary condition
is determined by a Kirchoff-like equation of the form
\begin{equation}
\sum_{i} S_{i}\frac{\partial\theta(\epsilon, x)}{\partial x} = 0
\label{boundequd}
\end{equation} 
\end{mathletters}
where $S_{i}$ denotes the cross section of the branch $i$ joining the
node \cite {golubov}.

Conceptually, at least, determining the resistance of any arbitrary sample is a
straightforward matter:  one solves the Usadel equation for the pair amplitude
$\theta (\epsilon, x)$ subject to the appropriate boundary conditions, then
substitutes the result into either Eq.\ (\ref{eq:IV}) or
(\ref{eq:RT}).  Practically, however, the Usadel equation needs to be solved
numerically.  This is not easy, particularly for complicated structures such as
our Andreev interferometers, and we make certain simplifying assumptions to make
the calculation tractable. 	Fig. 6(a) shows a schematic of sample B.  As we have
noted earlier, we assume that the N-S interface resistances are very small, so
that the loop resistance is zero, as it is shorted by the superconducting arm.  The
measured normal-metal resistance $R$ is then simply the sum of the two N side
branches, $R_{1}$ and $R_{2}$.  To determine the resistance of these structures in
the proximity effect regime, we need to solve the Usadel equation in the one
dimensional wires on either side of the loop.  Since the electron phase coherence
length $L_{\varphi}$ places an upper cutoff to the pair correlation in the normal
metal, we take the normal reserviors to be at a distance $L_{\varphi}$ from the
superconductor.  Finally, our calculations show that the effect of the voltage
probes on $\theta(\epsilon, x)$ in the side arms of the structure is very small,
and hence we ignore the effect of these probes.

Figure 6(b) shows the final N-S-N geometry that we simulate based on the procedure
developed in Refs. \cite {yip,nazarov2,golubov}.  If we consider the
superconductor to be at zero voltage, the energy $\epsilon$ of the quasiparticles
in each branch $i$ of the structure is related to the voltage drop $V_i$ between
the corresponding normal reservoir and the superconductor by
$\epsilon_i=eV_i$.  However, due to the four terminal nature of our measurements,
the voltage that is actually measured is the voltage $V_{2}$-$V_{1}$ at the voltage
probes.  Although the potential profile in the one dimensional normal wire between
the normal and superconducting reservoirs is not predicted to be linear \cite
{stoof}, the deviations from linearity are small enough that we can relate the
voltages measured at each probe to the voltage $V$ at the corresponding normal
reservoir by a linear scaling of the form
$V_{i}=V(L_{i}/L_{\varphi}$), where $L_{i}$ is the length of the arm from the
superconductor to the point at which the voltage probe joins the wire.
($L_{\varphi}$ is the effective distance to the corresponding normal reservoir.) 
Furthermore, to calculate the resistance of each branch, we need to use
Eq.\ (\ref{eq:RT}), but with the integral over the length $L$ restricted to 
$L_{i}$ for each branch, and the normal state resistance $R_{N}$ corresponding to
the normal state resistance
$R_{Ni}$ for each arm.  The total temperature dependent resistance of the sample
is then the sum of the resistances $R_{1}(T)$ and $R_{2}(T)$.

\subsection{Temperature dependence}

Figure 7(a) shows $R_{1}(T)$, $R_{2}(T)$, and the sum  $R_{1}(T)+R_{2}(T)$
calculated in the limit
$\Delta \gg E_{c}$ for the temperature regime below 1 K.  As expected, the major
contribution to the resistance comes from the longer (0.55
$\mu m$) side branch.   To calculate $\theta(\epsilon,x)$ we assume $\Delta$ is
much larger than all the energies integrated in Eq.\ (\ref{eq:RT}).  For the range
of the temperature of interest it is good enough to integrate up to $100 E_{c}$. 
Although this is larger than the actual gap ($\Delta = 32 E_{c}$ using BCS
theory),  changing $\Delta$ in this regime influences the
calculation by only a small amount in the region of interest.  Figure 7(b) shows the
result of the calculation for the total resistance using two different values of
$\Delta$.  In the regime $\Delta
\gg 100 E_{c}$, the curve recovers its normal state value faster at high temperature
than the curve obtained from the regime where $\Delta$ is close to $100 E_{c}$.  At
temperatures less than $5 E_{c}$, however, the two regimes do not show a significant
difference.  Therefore, in our simulations we assume $\Delta \gg 100E_{c}$.  

Regardless of the value of $\Delta$ assumed in the calculation, however, it is
clear that the proximity effect theory does not describe well the experimental
data in the high temperature ($T \leq T_{c}$) regime, as a comparison with Fig. 2
immediately shows.  This is because, near $T_{c}$, the contribution from the
superconducting transition in the superconducting arm of the loop (which we have
ignored so far) must be taken into account.  For our control pure Al wire, which
has a width of $\sim 0.15$
$\mu m$, the superconducting transition is fairly sharp, occurring within  a range
of 5-10 mK.  The resistance decrease near $T_{c}$ seen in Fig. 2, if it is indeed
due to the superconducting transition of Al, is much wider.  In quasi-one
dimensional pure superconducting wires, one mechanism that leads to a
finite resistance below the nominal superconducting transition is nucleation of
phase-slip centers.  The resistance broadening due to these phase-slip centers is
given by the Langer-Ambegaokar (LA) form \cite {tinkham}
\begin{equation} 
R = \frac {\alpha}{T}  exp \left[ -\beta \left( 1-\frac{T}{T_{c}}\right) ^{3/2}/T
\right]
\label{eq:LA}
\end{equation} 
where $\alpha$ is a parameter associated with the attempt frequency
for a phase slip event, and $\beta$ is a parameter related to the energy barrier
for a phase slip event.  For pure superconducting wires, $\beta$ is typically
very large ($\sim 10^{6}$ K) which confines the broadening of the
transition to a few millikelvin near $T_{c}$ \cite {tinkham}.  For our N-S
structures, since the experimentally measured transition is much broader,
$\beta$ is expected to be much smaller. 	

Figure 8 shows a comparison between the experimental data (triangles) and the
theoretical curve taking into account both the proximity effect and the influence
of phase slip centers.  The contribution from the proximity effect (dashed line)
was obtained as discussed above.  The influence of phase slip centers was
determined by fitting the difference between the experimental data and the
proximity effect contribution to the LA equation (\ref{eq:LA}), using $\alpha$ and
$\beta$ as fitting parameters.   The resulting contribution is shown as the dashed
line in Fig. 8, with the fitting parameters
$\alpha=0.17$ $\Omega$K and $\beta=9.93$ K.  While the contribution due to phase
slip centers is significant above 0.6 K, the contribution due to the proximity
effect shows a strong temperature dependence only below 0.4 K.  The sum of the two
contributions is shown as the solid line in the figure, and shows an excellent fit
to the experimental data.  We should remark that it is not clear that the theory
of phase slip centers should be applicable at all here, although our samples show
strong evidence that nonequilibrium phenomena are indeed important near the
transition \cite {chenjung}.  Nonetheless, our analysis does point to the fact
that the value of 
$T_{min}$  clearly depends on the interplay between the contribution due to the
proximity effect in the normal wire and the decrease in resistance
of the superconductor near $T_c$, and consequently, the measured $T_{min}$ will not
be simply related to the reentrance effect.

\subsection{Voltage dependence}

A similar analysis can be used to calculate the differential resistance
$dV/dI=R(V)$ as a function of voltage $V$.  To calculate $R(V)$, one notes the
temperature kernel in the integrand of Eq.\ (\ref{eq:IV}) becomes a step
function at $T = 0$ with the discontinuity centered at $\epsilon=eV$, i.e., the
contribution to the total current $I$ comes only from
$\left| \epsilon \right| < eV$ \cite {yip}.  The conductance $dI/dV$ thus contains
a $\delta$-function centered at
$\left| \epsilon \right | = eV$.  This results in a simple formula at $T=0$
\cite {yip}:
\begin{equation} R(V) =
R_{N}\left(\frac{1}{L}\int_{0}^{L}sech^{2}\left[Im\theta(\epsilon,x)\right]dx\right)_{\epsilon
= eV}
\label{eq:RV}
\end{equation}

This formula is applicable to each branch in Fig. 6(b).   In order to calculate
the total resistance $R(V)$, however, one must take into account that the voltage
$V$ across the entire sample is the sum of the voltages $V_{1}$ and $V_{2}$  across
each individual branch, subject to the condition that the current through both
branches is the same.  Figure 9 shows the experimental data and the calculated
curve based on the procedure outlined above.  The agreement between theory and
experiment is clearly not satisfactory.  The minimum in the experimental
resistance occurs at a voltage of $\sim$ 7.25
$\mu V$, while the theoretical curve shows a minimum at $\sim$ 25 $\mu V$.  Using
the same rationale as for the temperature dependent resistance, we attempt to add
the contribution from phase slip centers using the theory of LA \cite {tinkham}. 
However, with the same parameters obtained in Fig. 8, the curve obtained does
not agree with the experimental data, primarily due to the strong exponential
dependence of the theoretical result on the measuring current \cite{tinkham}.  There
are a number of possible reasons for this disagreement.  First, although we think
it is essential that nonequilibrium effects need to be taken into account in
discussing the voltage dependence, the theory of LA which was developed to discuss
phase slip centers in pure superconductors may not adequately describe
nonequilibrium phenomena in N-S devices.  Second, our calculations assume perfect N
and S reservoirs, which are not realized in the experiment.  Although this factor
might be expected to affect both the temperature and the voltage dependence,
indications are that the effect on the voltage dependence might be more
significant\cite {courtoispriv}.

\subsection{Magnetoresistance oscillations}	

The loop in our Andreev interferometers is essential for the observation of
magnetoresistance oscillations, and hence cannot be ignored in any calculation of
the magnetoresistance.  This complicates the calculation tremendously. 
Consequently, we have not attempted to numerically solve the quasiclassical
Green's function equations in the presence of a magnetic field.  However, the
qualitative behavior can be understood by drawing on our experience with other
quantum interference phenomena in doubly-connected geometries.  For the case of
weak localization in single normal metal rings, for example, the magnetoresistance
oscillates as a function of magnetic field with fundamental period $h/2e$ \cite
{washburn}.  The oscillations are suppressed exponentially with the phase
coherence length
$L_{\varphi}$,
$exp(-L/ L_{\varphi})$, where $L$ is the perimeter of the loop.  In our Andreev
interferometers, quantum coherence is maintained in the superconducting arms of
the loop.  In the normal arm, the oscillation amplitude is determined by the phase
coherence length $\xi_{N}$.  Since the normal arm is connected to the
superconductor on both sides, the suppression of the oscillation amplitude might
be expected to go as $\sim exp[-L/(2 \xi_{N})]$, where $L$ is now the length of
the normal arm.  This exponential dependence is what we indeed observe in sample
A, and also in sample B at higher temperatures.  The reentrance effect we see in
the amplitude of the magnetoresistance oscillations in sample B is an indication
that these oscillations are dependent on the enhancement of the diffusion
coefficient in the normal arm of the loop.    

\section{Conclusion}	 

In conclusion,  we have investigated the reentrance effect in two mesoscopic N-S
hybrid loops with different interface transparencies.  The low transmittance
sample showed no reentrant behavior, consistent with the fact that the relevant
energy and temperature scales were shifted to values below our measurement range. 
The high transmittance sample, on the other hand, showed reentrant behavior in
$R(T)$, $dV/dI(V)$, and amplitude of magnetoresistance oscillations, due to the
long range coherence of the electron-hole pairs induced by Andreev reflection at
the N-S interfaces.  A quantitative understanding of the experimental results
cannot be obtained from the quasiclassical Green's function theory of reentrance
alone.  For a more complete quantitative understanding of the properties of such
N-S devices, we believe it is essential to understand the effect of nonequilibrium
phenomena on the transport properties, particularly at high bias voltages or near
the transition temperature of the superconductor. 

\acknowledgments

We acknowledge useful discussions with Sungkit Yip and  John Ketterson, and thank
Michael Black for a critical reading of the manuscript.  Work at Northwestern was
supported by the NSF under DMR-9357506, and by the David and Lucile Packard
Foundation.

\begin{figure}
\caption{(a) Scanning electron micrograph of sample A.  The additional gate
electrode was kept grounded and not used in these measurements.  (b) Sample
schematics for the two samples.  The dimensions are indicated in $\mu$m. The leads
used to applied ac currents and measure the voltages are also shown in the
schematics.  For the $dV/dI$ measurements, an additional dc current is applied
through I+/-.}
\end{figure}

\begin{figure}
\caption{The normalized resistance $R/R_{N}$ for samples A and B as a function of
temperature
$T$. 
$R_{N}$=67.5 $\Omega$ and 10.3 $\Omega$  for samples A and B respectively.}
\end{figure}

\begin{figure}
\caption{(a), (b) are the magnetoresistance curves $R(H)$ for sample A and B
respectively.  The small offset of $H$ in (a) is due to the residual flux trapped
in the superconducting magnet. In (a) the curves for $T$=101 mK, 203 mK, 400 mK
and 1.07 K are shifted up by 4 $\Omega$, 8 $\Omega$, 12 $\Omega$, and 16 $\Omega$
respectively.  In (b) the curves for $T$ = 97 mK, 199 mK, 491 mK and 600 mK are
shifted up by 0.2
$\Omega$, 0.4 $\Omega$, 0.7 $\Omega$ and 0.8 $\Omega$ respectively.   (c)
Normalized amplitude of the Fourier transform of (a) and (b) as a function of
temperature.  The field range is +/- 25 mT and +/- 20 mT and the normalization
constant is 0.982 $\Omega$ and 0.019 $\Omega$ for samples A and B respectively.
The solid lines represent fits to the form $a exp(-b T^{1/2})$ at higher
temperatures, with a = 3.1, 2.7 and b = 3.4, 2.2 for samples A and B
respectively.  For comparison, we also show a best fit to a power law of the form
$a/T$ as used by Courtois \emph{et al.} \protect \cite {courtois}, with the values
a=0.128, 0.238 for samples A and B respectively.}
\end{figure}

\begin{figure}
\caption{Normalized $dV/dI$ as a function of dc voltage $V$ at $T$=30 mK for
sample A (solid curve) and B (dotted curve).  The voltage  is obtained by
integrating
$dV/dI$ vs.
$I_{dc}$.  ac and dc currents are applied through I+/- shown in Fig. 1(b). 
$(dV/dI)_{N}$ is 67.5
$\Omega$ and 10.3 $\Omega$ for samples A and B respectively.  The dashed curve
shows $dV/dI$ for sample A at a magnetic field of 225 gauss.}
\end{figure}

\begin{figure}
\caption{$R(T)$ measured for (a) sample A, and (b) sample B at various values of
integral and half-integral flux quanta $h/2e$ through the loop.  Closed symbols,
integral flux quanta, open symbols, half-integral flux quanta.  The temperature
dependent curves were obtained from data similar to that of Fig. 3.}
\end{figure}

\begin{figure}
\caption{Schematic of the simulated model.  (a) Actual sample. (b) Geometry
used in the calculation.}
\end{figure}

\begin{figure}
\caption{Simulation of the resistance as a function of temperature based on the
model in Fig. 6.  (a) Resistance as a function of temperature for $R_{1}$, $R_{2}$
and
$R=R_{1}+R_{2}$, assuming $\Delta \gg E_{c}$.   (b) Total resistance $R$ as a
function of $T$ for different $\Delta$'s.  $\Delta = 200 E_{c}$, $10^{4} E_{c}$ for
the solid curve and dashed curve respectively.}
\end{figure}

\begin{figure}
\caption{Theoretical fit to $R(T)$ combining the predictions of the quasiclassical
Green's function theory with the phase slip model of Langer and Ambegaokar
\protect \cite{tinkham}, as discussed in the text. The parameters used in the fit
(referring to Eq.\ (\ref{eq:LA})) are $\alpha=0.17$ $\Omega$K, and
$\beta=9.93$ K,
$T_{c}$ = 1.2 K.  The dotted curve which represents the prediction of the
quasiclassical Green's function theory, is shifted down by 0.04
$\Omega$ and the dashed curve, which represents the predictions of the phase slip
model of LA is shifted up by 0.45 $\Omega$ for clarity.}
\end{figure}

\begin{figure}
\caption{Comparison of the theoretical calculation (solid curve) based on
the theory of quasiclassical Green's functions with the measured $dV/dI(V)$
(triangles).  The dashed curve shows the effect of adding the predictions of the
LA theory using the same parameters as in Fig. 8.}
\end{figure}


\begin{references}
\bibitem{lphinote}$L_{\varphi}$ is estimated from weak localization measurements
on co-evaporated normal metal control wires.  We believe $L_{\varphi}$ for the Ag
wire is very short due to the bake required for the second layer of the
lithography.
\end{references}
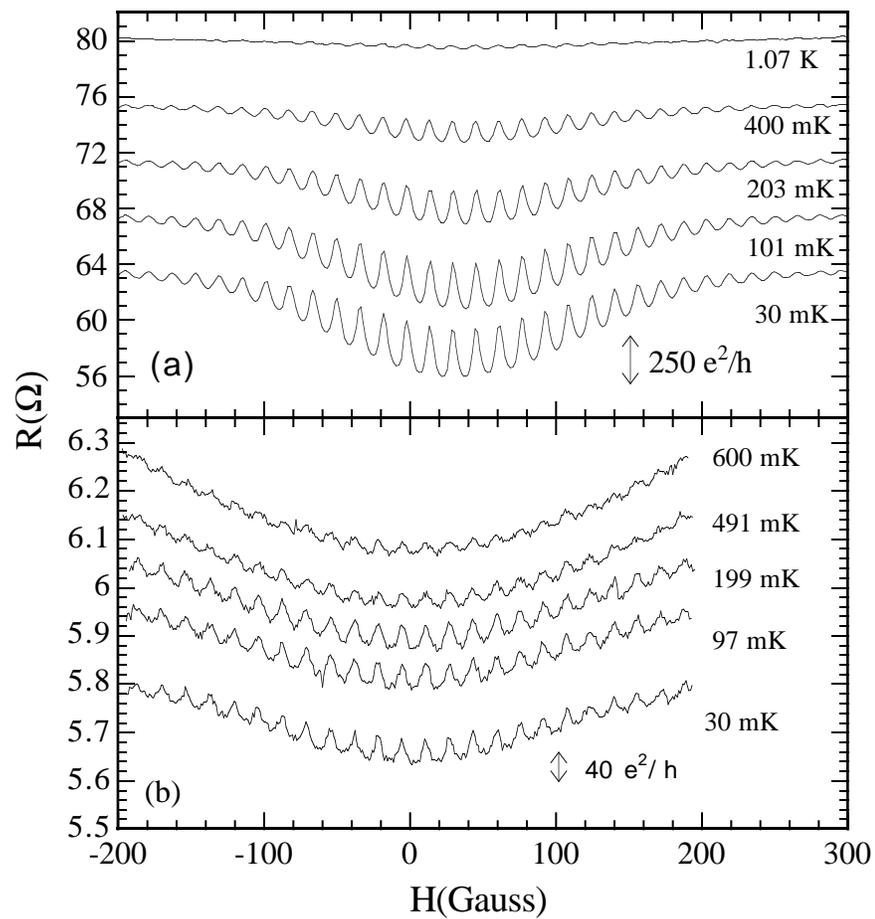
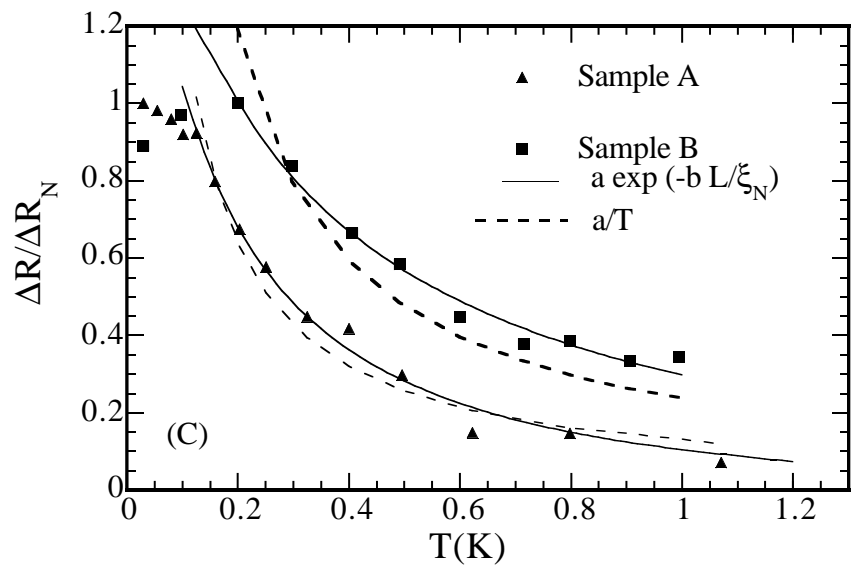
\end{document}